# OBSERVATION OF A H- BEAM ION INSTABILITY

Milorad Popovic and Todd Sullivan

Fermi National Accelerator Laboratory[1]

Batavia, IL 60510, USA

*Abstract*

We report the results of observations of H- beam instabilities at the Fermilab Linac. By intentionally creating "high" background pressure with different gases in the 750 keV transport line we observed coherent transverse beam oscillations. The minimal pulse length required to observe oscillations and the frequency of oscillations are functions of pressure and mass of the background gas. The oscillations are present in both transverse planes and very quickly reach saturation in amplitude growth. The observed characteristics of beam oscillations are in quantitative agreement with "fast beam-ion instability" described by Raubenheimer and Zimmermann[1]. Effects described here are occur far from the normal operating range of the Fermilab Linac but may be important for many future high intensity accelerators.

## 1 INTRODUCTION

In 1985 a BPM system was introduced in the Fermilab Linac[2] and fast H- beam transverse oscillations were noticed when the pressure of the 750 keV line was degraded by turning off the large ion pump near the H- source. Recently we have revisited this phenomena in the light of renewed interest in this type of beam instability. In many future rings, this transient instability is predicted to have very fast growth rates, much faster than the damping rates of existing and proposed transverse feedback systems, and thus is a potential limitation. The instability described in this paper is caused by residual gas ions. Charged particle beams, traversing a beam line or circulating in a storage ring, ionize the residual gas and generate free electrons and ions. The instability mechanism is the same in the beam line and storage rings assuming that ions are not trapped turn by turn in the rings. The ions generated by the head portion of the beam pulse oscillate in the transverse direction causing a growth of the initial perturbation of the beam. In our case, ions of the background gas are trapped and focused by H- beam. They start to oscillate and create transverse deformation of the H- beam. The model employed by Raubenheimer and Zimmermann is in quantitative agreement with our observations. In this model all ions oscillate with the same frequency, the frequency of small-amplitude oscillations of the centroid in the potential well of the beam. In our experiment we see a frequency spread of oscillations which increases with pressure.

## 2 EXPERIMETAL SETUP

During the experiment, a 750keV H- DC beam was transported along a 10 meter long transport line to the buncher cavity and Drift Tube Linac, see figure 1. The beam size along the line does not change significantly and averages about 2 cm in diameter. The peak current of the beam was 65 mA at a starting background pressure of 1.2e-6 Torr. The background gas was mostly hydrogen gas from the Ion source. The pulse length of 35 us is created using a chopper which is at the beginning of the line. Vacuum in the transfer line was maintained using a turbo pump near the Chopper and another by Tank#1. The pressure in the line was measured using an Ion Gauge near the gas bleeding valve that was used for introducing different gases and creating different pressures in the line.

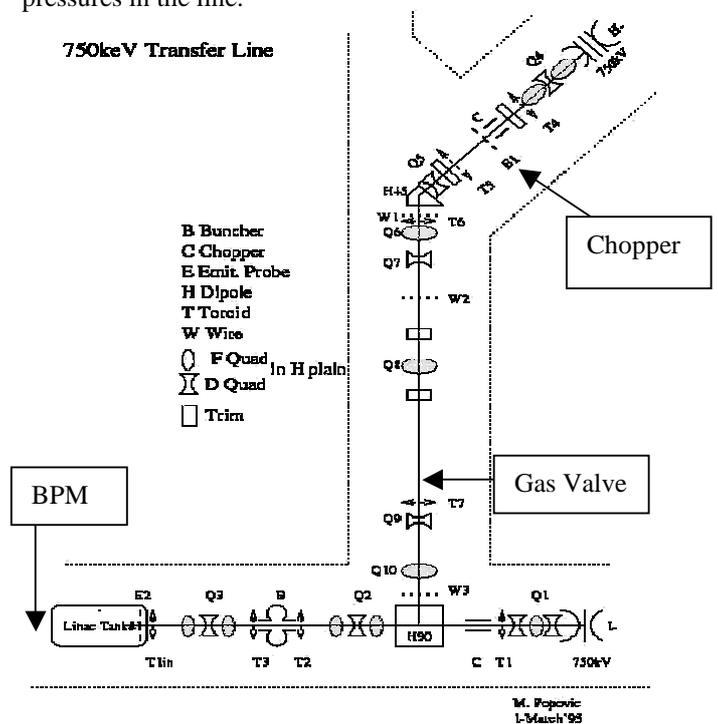

**Figure 1**

Beam current was measured at the entrance to Tank#1 and at the exit of Tank#2. The beam signal on the BPM after Tank#2 was used to observe beam centred position

[1] This work is supported by U.S. Dept. Of Energy through the University Research Association under contract DE-AC35-89ER40486

during the pulse. The signals were recorded using a LeCroy scope. For measurements of the beam oscillation frequency, FFT signals were averaged over many beam pulses. To avoid any frequency signal not related with beam oscillations only the last 30 us of the beam pulse was Fourier analysed. There was no noticeable difference between horizontal and vertical planes, so all data was taken looking at the horizontal plane only.

## 2 RESULTS

Using a bleeding valve in the middle of the transfer line measurements were repeated for several different gases. We used hydrogen, helium, argon and krypton as the background gas. With the bleeding valve we where able to
have fine control of the pressure in the line.

linac, see figure 2. It is known that transmission through the linac is very dependent on the quad settings in the line. We have not measured beam profiles in the line but know that the Buncher is an aperture restriction in the line, and the high transmission is achieved only if the beam has a waist at the Buncher position. We can say with some degree of confidence that the high background gas pressure did not change the beam profile in the line. Under normal operation, the pressure in the line is $2.4 \times 10^{-6}$ and in the experiment the highest pressure was $1.0 \times 10^{-4}$.

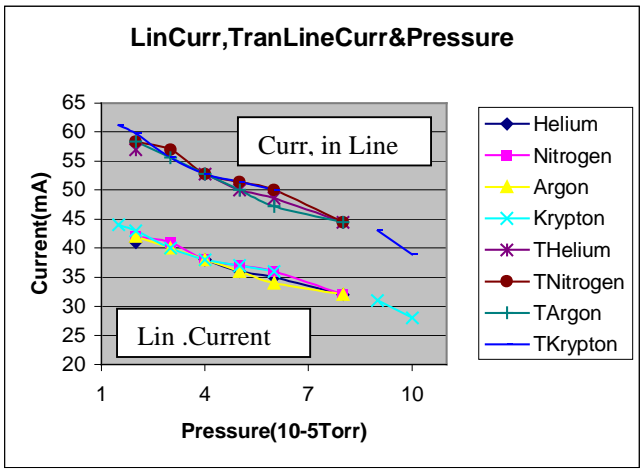

Figure 2

Two toroids were used to record the beam current in the transfer line to insure that the current was constant in the line for the whole range of pressure and gases.

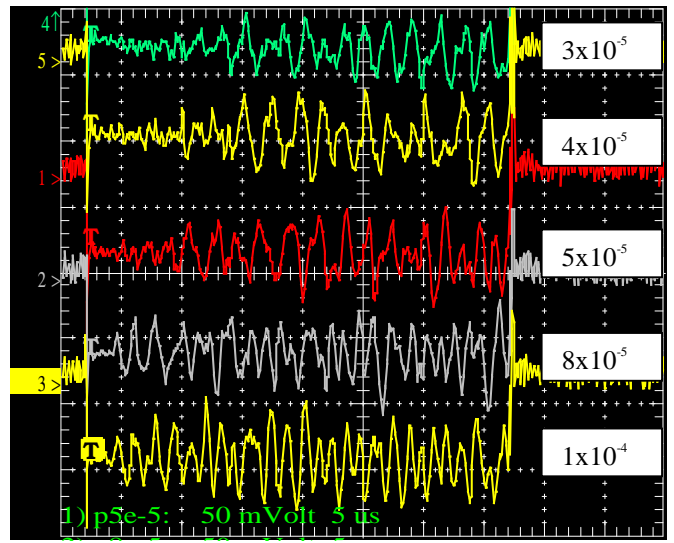

Figure 4

The lower trace in figure 3 is a beam position signal from the BPM at the entrance to Tank#3 under normal operation with pressure in the line of $2.4 \times 10^{-6}$ Torr. Small fluctuation of the signal are result of the noise in the beam and pick-up. Upper trace is the same signal with a pressure in the line of $4.8 \times 10^{-5}$ Torr. It is clear that

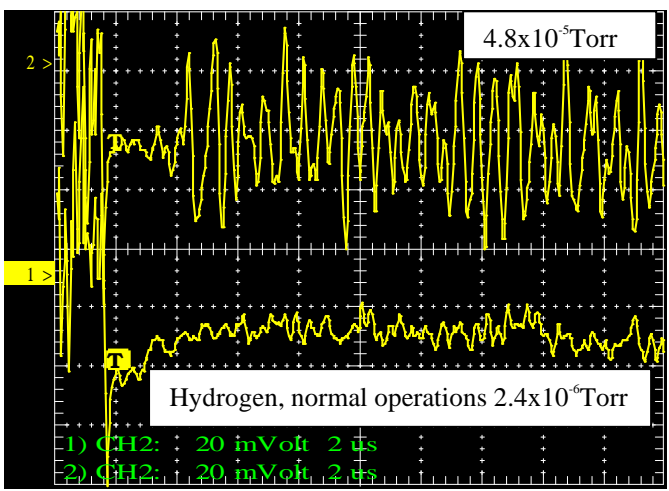

Figure 3

As a way of monitoring that focusing properties of the line did not change, we recorded beam current at end of oscillations start about 1.2 µs after the start of beam, develop very quickly and saturate after one or two full oscillations. The time for oscillations to develop and the frequency of oscillations depend on the pressure and type of background gas. Figure 4 shows scope traces of the beam position for different gas pressures in Torr, when the background gas is argon. To measure the frequency of oscillations we used an FFT option built into the LeCroy scope.

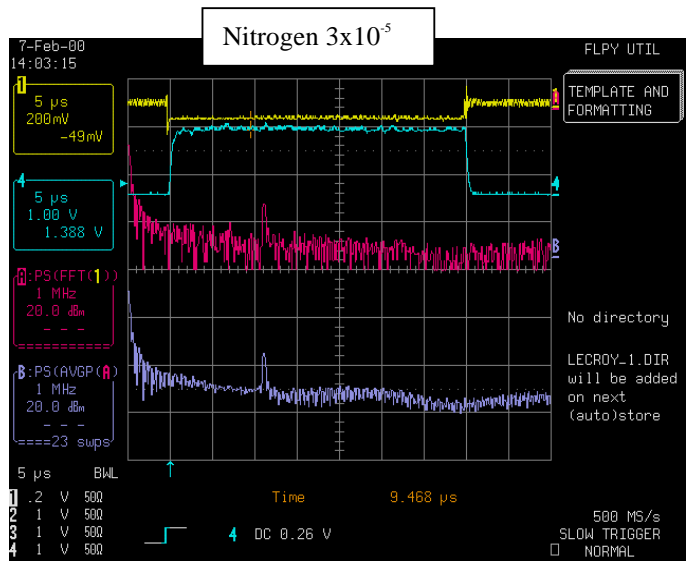

Figure 5

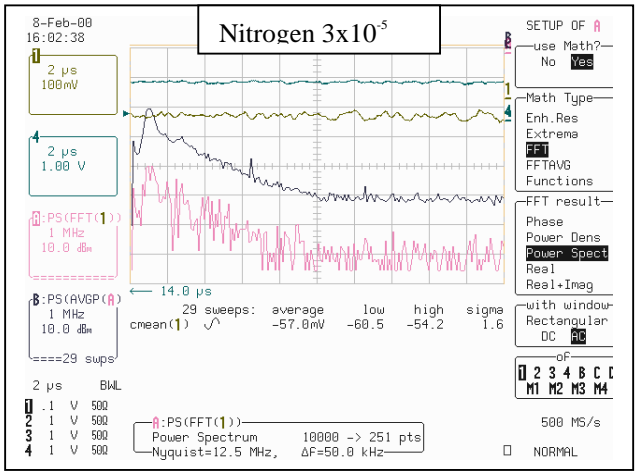

Figure 5

Figure 5 shows four traces: beam position, beam intensity, FFT of beam position signal and the average of the last 23 FFT beam signals. To exclude no-beam related signals as in Figure 5 that come from noise on the pick-up, we used only the last 30 us of the beam signal. For low gas pressure we see a relatively sharp frequency signal in the range of 0.5 MHz. As pressure is increased the frequency signal is broadened and moves toward higher frequency with the peak at 1.1 MHz for all gases.

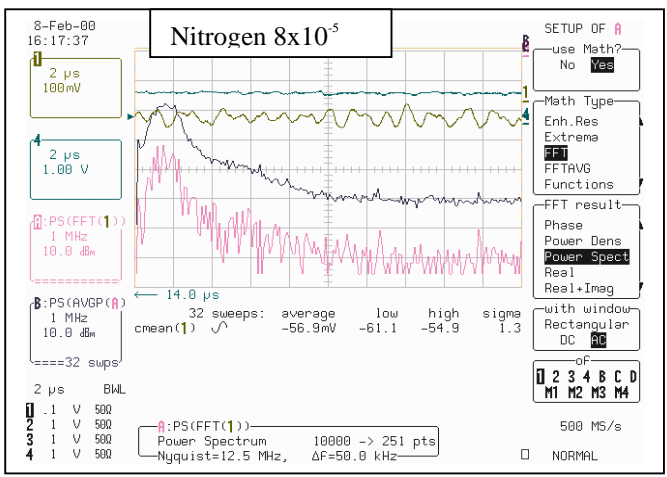

Figure 6

## 4 CONCLUSIONS

We have intentionally created coherent transverse beam oscillations. The oscillations have the following characteristics:
- The time to develop oscillations strongly depends on the pressure but not much to the gas species,
- It takes only a few oscillations for the instability to fully develop and saturate.
- At low gas pressure, the oscillation frequency is ~0.5MHz .
- At higher gas pressure, the oscillation frequency peaks at ~1.1MHz for all gas species.

## REFERENCES

[1] E. McCrory, G. Lee and R. Webber, "Observation of Transverse Instabilities in FNAL 200MeV Linac", 1988 Linear Accelerator Conf. CEBAF-89-001, pp 182-184.
[2] T.O.Raubenheimer and F. Zimmermann, Phys. Rev. B52, 5487(1995).